%% file: QCD-10-025_temp.tex
\pdfoutput=1

\documentclass[11pt,twoside,a4paper,cmspaper,final,collab]{cms-tdr}
\def\svnVersion{49526M}\def\cmsCernNo{2011-042}\def\cmsCernDate{\today}\def\cmsMessage{Submitted to Physics Letters B}

\begin{document}\cmsNoteHeader{QCD-10-025}

\hyphenation{had-ron-i-za-tion}
\hyphenation{cal-or-i-me-ter}
\hyphenation{de-vices}
\RCS$Revision: 49369 $
\RCS$HeadURL: svn+ssh://alverson@svn.cern.ch/reps/tdr2/papers/QCD-10-025/trunk/QCD-10-025.tex $
\RCS$Id: QCD-10-025.tex 49369 2011-04-08 09:07:25Z kkousour $
\cmsNoteHeader{QCD-10-025} 
\title{Measurement of the differential dijet production cross section in proton-proton collisions at $\sqrt{s}=7\TeV$}

\author[cern]{The CMS Collaboration}
\date{\today}

\abstract{
A measurement of the double-differential inclusive dijet production cross section in proton-proton collisions at $\sqrt{s}=7\TeV$ is presented as a function of the dijet invariant mass and jet rapidity. The data correspond to an integrated luminosity of $36\pbinv$, recorded with the CMS detector at the LHC. The measurement covers the dijet mass range $0.2\TeV$ to $3.5\TeV$ and jet rapidities up to $|y|=2.5$. It is found to be in good agreement with next-to-leading-order QCD predictions.
}

\hypersetup{%
pdfauthor={CMS Collaboration},%
pdftitle={Measurement of the differential dijet production cross section in proton-proton collisions at sqrt(s)=7 TeV},%
pdfsubject={CMS},%
pdfkeywords={CMS, physics, QCD, jets}}

\maketitle

In Quantum Chromodynamics (QCD), events with two high transverse momentum jets (dijets) arise in proton-proton collisions from parton-parton scattering, where the outgoing scattered partons manifest themselves as hadronic jets. The invariant mass $M_\text{JJ}$ of the two jets is related to the proton momentum fractions $x_{1,2}$ carried by the scattering partons: $M_\text{JJ}^2=x_1\cdot x_2\cdot s$, where $\sqrt{s}$ is the centre-of-mass energy of the colliding protons. The dijet cross section as a function of $M_\text{JJ}$ can be precisely calculated in perturbative QCD and allows also a sensitive search for physics beyond the Standard Model, such as dijet resonances or contact interactions. In this Letter, the measurement of the double-differential inclusive dijet production cross section (p + p $\rightarrow$ jet + jet + X) is reported as a function of the dijet invariant mass and jet rapidity at $\sqrt{s}=7\TeV$. The data were collected with the Compact Muon Solenoid (CMS) detector at the CERN Large Hadron Collider (LHC) during the 2010 run and correspond to an integrated luminosity of $36\pbinv$. The measured cross section is compared to the QCD predictions in an unexplored kinematic region, beyond the reach of previous measurements~\cite{D0,CDF,ATLAS}. The parton momentum fractions probed in this measurement correspond to $8\cdot 10^{-4}\leq x_1\cdot x_2\leq 0.25$. Dedicated searches for dijet resonances and contact interactions with the CMS detector have been reported elsewhere~\cite{CMSSearchMass,CMSSearchRatio,CMSSearchAngular}.

The CMS coordinate system has its origin at the centre of the detector, with the $z$-axis pointing along the direction of the counterclockwise beam. The azimuthal angle is denoted as $\phi$, the polar angle as $\theta$, and the pseudorapidity is defined as $\eta=-\ln\left[\tan\left(\theta/2\right)\right]$. The central feature of the CMS detector is a superconducting solenoid, of 6\,m internal diameter, that produces an axial magnetic field of 3.8\,T. Within the field volume are the silicon pixel and strip tracker, a lead-tungstate crystal electromagnetic calorimeter (ECAL) and a brass/scintillator hadronic calorimeter. Outside the field volume, in the forward region ($3 < |\eta| < 5$), is an iron/quartz-fiber hadronic calorimeter. Muons are measured in gas detectors embedded in the steel return yoke outside the solenoid, in the pseudorapidity range $|\eta| < 2.4$. A detailed description of the CMS experiment can be found in Ref.~\cite{CMS}.

Jets are reconstructed using the anti-$k_T$ clustering algorithm~\cite{AKT} with size parameter $R=0.7$. The clustering is performed using four-momentum summation, where the chosen size parameter allows for the capture of most of the parton shower and improves the dijet mass resolution with respect to smaller sizes. The rapidity $y$ and the transverse momentum \pt of a jet with energy $E$ and momentum $\vec{p}=(p_x,p_y,p_z)$ are defined as $y=\frac{1}{2}\ln\left(\frac{E+p_z}{E-p_z}\right)$ and $\pt=\sqrt{p_x^2+p_y^2}$, respectively. The inputs to the jet clustering algorithm are the four-momentum vectors of the reconstructed particles. Each such particle is reconstructed with the particle-flow technique~\cite{PFLOW} which combines the information from several subdetectors. The resulting jets require an additional energy correction to take into account the non-linear and non-uniform response of the CMS calorimetric system to the neutral-hadron component of the jet (the momentum of charged hadrons and photons is measured accurately by the tracker and the ECAL, respectively). The jet-energy corrections are derived using simulated events, generated by {\sc pythia6.4.22} ({\sc pythia6})~\cite{PYTHIA} and processed through the CMS detector simulation based on {\sc geant4}~\cite{GEANT4}, and \textit{in situ} measurements with dijet and photon+jet events~\cite{JME-10-010}. An offset correction is also applied to take into account the extra energy clustered in jets due to additional proton-proton interactions within the same bunch crossing (pile-up). The jet-energy correction depends on the $\eta$ and \pt of the jet, and is applied as a multiplicative factor to the jet four-momentum vector. The multiplicative factor is in general smaller than 1.2. For a jet $\pt=100\GeV$ the typical factor is 1.1, decreasing towards 1.0 with increasing \pt. The dijet mass is calculated from the corrected four-momentum vectors of the two jets with the highest \pt (leading jets): $M_\text{JJ}=\sqrt{\left(E_1+E_2\right)^2-\left(\vec{p}_1+\vec{p}_2\right)^2}$. The relative dijet-mass resolution, estimated from the simulation, ranges from 7\% at $M_\text{JJ}=0.2\TeV$ to $3\%$ at $M_\text{JJ}=3\TeV$.

The data samples used for this measurement were collected with single-jet high level triggers (HLT)~\cite{HLT} which required at
least one jet in the event to satisfy the condition $\pt > 30$, $50$, $70$, $100$ and $140$ \GeV, respectively, in uncorrected jet transverse momentum. The lower-\pt triggers were prescaled and the corresponding integrated luminosity of each trigger sample, $\mathcal{L}_{\text{eff}}$, is listed in Table~\ref{tab:lumi}. In the offline analysis, events are further required to have at least one well reconstructed proton-proton interaction vertex~\cite{TRK-10-005} and at least two reconstructed particle-flow jets with ${\pt}_1>60\GeV$ and ${\pt}_2>30\GeV$ (corrected). In order to suppress nonphysical events, the two leading jets must satisfy loose identification criteria: each jet should contain at least two particles, one of which is a charged hadron. Furthermore, the jet energy fraction carried by neutral hadrons and photons should be less than 99\%. If either of the leading jets fails the identification criteria, the event is discarded. The measurement is performed in five rapidity regions, defined by the maximum absolute rapidity $|y|_\text{max}=\text{max}\left(|y_1|,|y_2|\right)$ of the two leading jets in the event. The use of the variable $|y|_\text{max}$ divides the phase space of the dijet system into exclusive rapidity bins, which correspond to different scattering angles at the centre-of-mass frame. Low values of $|y|_\text{max}$ probe the large-angle scattering (\textit{s channel}), while large values of $|y|_\text{max}$ probe the small-angle scattering (\textit{t channel}). For the construction of the invariant mass spectrum, each dijet-mass bin is populated by events collected only with the fully efficient trigger with the highest threshold. The efficiency for each trigger path was measured using events collected with a lower threshold single-jet trigger and cross-checked with events collected with single-muon triggers.

\begin{table}[htbH]
   \begin{center}
      \caption{The integrated luminosity for each of the minimum (uncorrected) jet-\pt data samples. \label{tab:lumi}}
      \begin{tabular}{|c|c|c|c|c|c|}
         \hline
         Minimum jet \pt (\!\GeV) & 30 & 50 & 70 & 100 & 140 \\
         \hline
         $\mathcal{L}_{\text{eff}} (\!\pbinv)$ & 0.32 & 3.2 & 8.6 & 19 & 36 \\
         \hline
      \end{tabular}
   \end{center}
\end{table}

\begin{figure}[hbtp]
   \begin{center}
      \includegraphics[width=0.95\columnwidth]{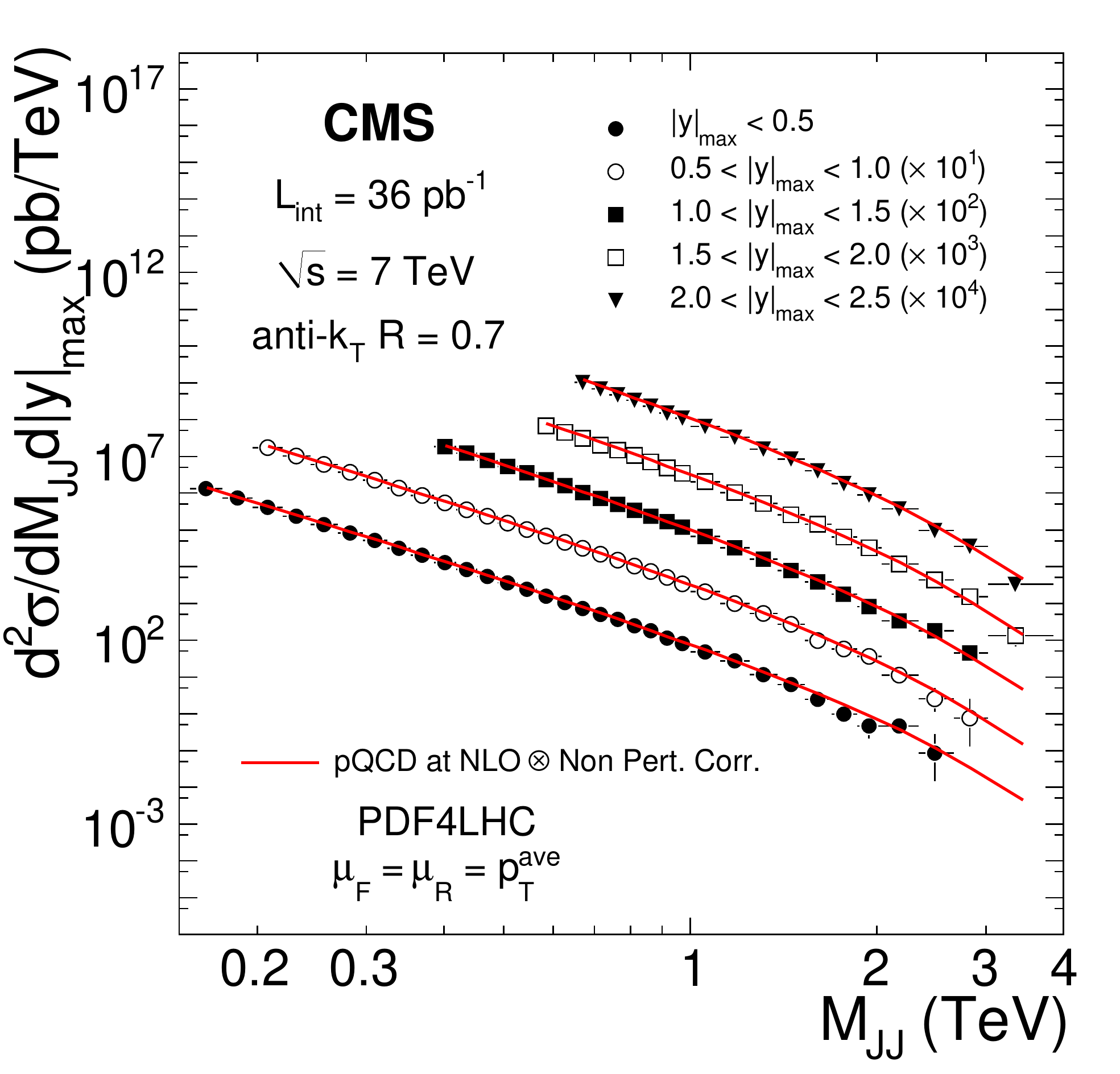}
      \caption{Measured double-differential dijet production cross sections (points), scaled by the factors shown in the figure, as a function of the dijet invariant mass, in bins of the variable $|y|_\text{max}$, compared to the theoretical predictions (curves). The horizontal error bars represent the bin widths, while the vertical error bars represent the statistical uncertainties of the data.}
      \label{fig:xsec}
   \end{center}
\end{figure}

The double-differential cross section is defined as

\begin{equation}\label{master_formula}
\frac{\text{d}^2\sigma}{\text{d}M_\text{JJ}\text{d}|y|_\text{max}}=\frac{\mathcal{C}}{\epsilon\mathcal{L}_{\text{eff}}}\frac{N}{\Delta M_\text{JJ}\Delta |y|_\text{max}}\text{,}
\end{equation}

where $N$ is the number of events in the bin, $\mathcal{L}_{\text{eff}}$ is the integrated luminosity of the data sample from which the events are taken, $\mathcal{C}$ is a correction factor for bin-to-bin migration, $\epsilon$ is the product of the trigger and event selection efficiencies (greater than $99\%$), and $\Delta M_\text{JJ}$ and $\Delta |y|_\text{max}$ are the invariant mass and rapidity bin widths, respectively. The width of the mass bins is progressively increased, proportional to the mass resolution. The correction factor $\mathcal{C}$ is taken from the simulation, as follows. Jets reconstructed with the same clustering algorithm from generated particles are smeared according to the simulated energy resolution and the correction factor $\mathcal{C}$ is defined as the ratio of the generated over the smeared numbers of events in a given dijet-mass bin. It ranges between 0.95 and 0.98, depending on the dijet mass and the rapidity region. Figure~\ref{fig:xsec} shows the double-differential cross section as a function of the dijet mass in different bins of $|y|_\text{max}$. The exact mass ranges and the cross-section values are reported in Tables~\ref{tab:ybin0}--\ref{tab:ybin4}. The quoted reference mass for each bin is the mass value $m_0$ that satisfies the equation $f(m_0)(m_2-m_1)=\int_{m_1}^{m_2}{f(m)\,dm}$, where $m_1,\,m_2$ are the bin boundaries and $f(m)=A\cdot\left(m/\sqrt{s}\right)^{-a}\cdot\left(1-m/\sqrt{s}\right)^b$, with parameters obtained from a fit to the mass spectrum. The definition of the reference mass follows the approach described in~\cite{POINT}.

The systematic uncertainty on the measured cross section is asymmetric and dominated by the uncertainty on the jet-energy scale. The latter varies between $3\%$ and $5\%$~\cite{JME-10-010} and introduces a $15\%$ ($60\%$) uncertainty on the cross section at $M_\text{JJ}=0.2\TeV$ ($3\TeV$). The uncertainty on the integrated luminosity is estimated to be 4\%~\cite{LUMI} and propagates directly to the cross section. The jet-energy resolution uncertainty of $10\%$~\cite{JME-10-014} propagates to the dijet mass resolution, which affects the unsmearing correction, introducing a $1\%$ uncertainty on the cross section. Other sources of experimental uncertainty, such as the jet angular resolution and the Monte Carlo \pt spectrum used to calculate the smearing effect, introduce negligible uncertainties on the cross section. The quoted experimental systematic uncertainties of the individual dijet-mass bins are almost $100\%$ correlated.

The theoretical prediction for the double-differential cross sections consists of a next-to-leading-order QCD calculation and a nonperturbative correction to account for the multiparton interactions (MPI) and hadronisation effects. The NLO calculations are done using the NLOJet++ program (v2.0.1)~\cite{NLO} within the framework of the fastNLO package (v1.4)~\cite{fastNLO} at renormalization and factorization scales ($\mu_R$ and $\mu_F$) equal to the average transverse momentum $\pt^\text{ave}$ of the two jets. The NLO calculation is performed using the CT10~\cite{CT10}, MSTW2008NLO~\cite{MSTW} and NNPDF2.0~\cite{NNPDF} parton distribution functions (PDF) at the corresponding three default values of the strong coupling constant $\alpS(M_Z)= 0.1180, 0.1202$ and $0.1190$, respectively, recommended by the PDF4LHC working group~\cite{PDF4LHC}. The central value of the NLO calculation is taken as the average of the minimum and the maximum values predicted by the envelope of the $68\%$ confidence level uncertainty of the three PDF. The non-perturbative effects are estimated from the simulation, using the event generators {\sc pythia6} (tunes D6T~\cite{D6T} and Z2~\cite{Z2}) and {\sc herwig++ 2.4.2}~\cite{HERWIG}. The non-perturbative correction is defined as the ratio of the cross section predicted with the nominal generator settings divided by the cross section predicted with the MPI and hadronisation switched off. The central value of the non-perturbative correction is calculated from the average of the three models considered, and ranges from $30\%$ at the lowest dijet mass value in each rapidity region, to 5\% at $M_\text{JJ}=3\TeV$. The PDF variation introduces a $5\%$ ($30\%$) uncertainty on the theoretical prediction at a dijet mass of $0.2\TeV$ ($3\TeV$), while the variation of $\alpS(M_Z)$ by 0.002 introduces an additional $2$--$4\%$ uncertainty. The renormalization and factorization scale uncertainty is estimated as the maximum deviation at the six points ($\mu_F/\pt^\text{ave},\mu_R/\pt^\text{ave}) =(0.5,0.5),(2,2),(1,0.5),(1,2),(0.5,1),(2,1)$, introducing a $+2\%$ ($+8\%$), $-5\%$ ($-13\%$) uncertainty at $M_\text{JJ}=0.2\TeV$ ($3\TeV$) in the central rapidity bin ($|y|_\text{max}<0.5$), and $+2\%$ ($+5\%$), $-10\%$ ($-32\%$) uncertainty at $M_\text{JJ}=0.7\TeV$ ($3\TeV$) in the outermost rapidity bin ($2.0<|y|_\text{max}<2.5$). An additional uncertainty of $15\%$ ($2\%$) at $M_\text{JJ}=0.2\TeV$ ($3\TeV$) is caused by the non-perturbative correction. Overall, the PDF uncertainty dominates at high masses, while the non-perturbative correction uncertainty is dominant at low masses.

\begin{figure*}[hbtp]
   \begin{center}
      \includegraphics[width=0.95\textwidth]{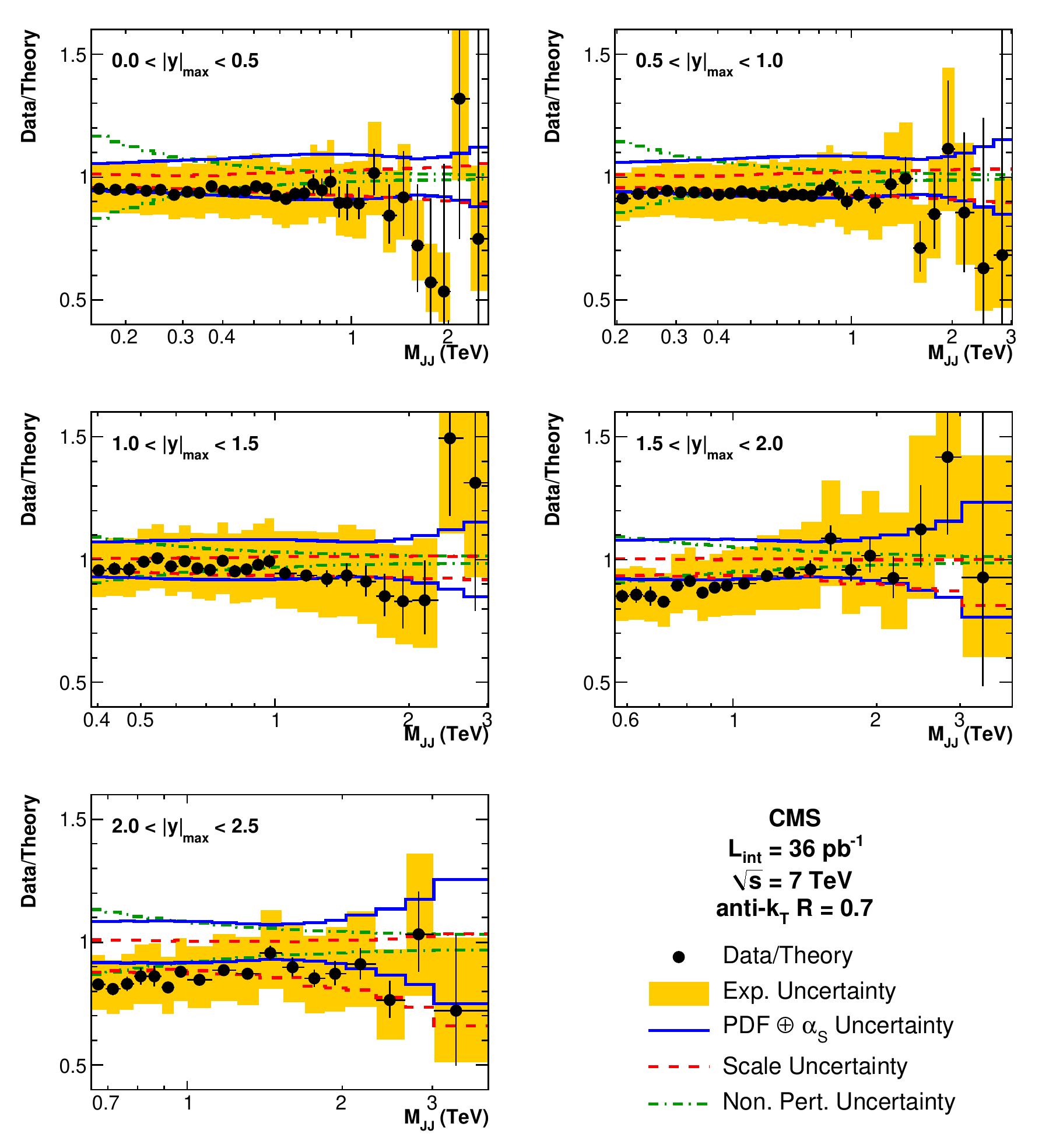}
      \caption{Ratio of the measured double-differential dijet production cross section over the theoretical prediction in different rapidity bins. The solid band represents the experimental systematic uncertainty and is centered around the points. The error bars on the points represent the statistical uncertainties. The theoretical uncertainties due to PDF and the strong coupling constant $\alpS(M_Z)$ (solid blue), renormalization and factorization scales (dashed red), and non-perturbative effects (dashed-dotted green) are shown as curves centered around unity.}
      \label{fig:dataOverNLO}
   \end{center}
\end{figure*}

Figure~\ref{fig:dataOverNLO} shows the comparison between the data and the theoretical prediction in the various bins of $|y|_\text{max}$. It also shows the components of the theoretical uncertainty. The agreement observed for the entire mass range and in all rapidity bins is good. The experimental uncertainty is comparable to the theoretical uncertainty and the data can be used to constrain the ingredients of the QCD calculations.

In summary, a measurement of the double-differential dijet cross section as a function of the dijet mass and $|y|_\text{max}$ has been presented. Using $36\pbinv$ of data from proton-proton collisions at $\sqrt{s}=7\TeV$ collected with the CMS detector, the measurement covers the dijet-mass range from $0.2\TeV$ to $3.5\TeV$ in five rapidity bins, up to $|y|_\text{max}=2.5$. The data are in good agreement with the theoretical prediction, showing that QCD accurately describes the parton-parton scattering in this kinematic region.

We wish to congratulate our colleagues in the CERN accelerator departments for the excellent performance of the LHC machine. We thank the technical and administrative staff at CERN and other CMS institutes, and acknowledge support from: FMSR (Austria); FNRS and FWO (Belgium); CNPq, CAPES, FAPERJ, and FAPESP (Brazil); MES (Bulgaria); CERN; CAS, MoST, and NSFC (China); COLCIENCIAS (Colombia); MSES (Croatia); RPF (Cyprus); Academy of Sciences and NICPB (Estonia); Academy of Finland, ME, and HIP (Finland); CEA and CNRS/ IN2P3 (France); BMBF, DFG, and HGF (Germany); GSRT (Greece); OTKA and NKTH (Hungary); DAE and DST (India); IPM (Iran); SFI (Ireland); INFN (Italy); NRF and WCU (Korea); LAS (Lithuania); CINVESTAV, CONACYT, SEP, and UASLP-FAI (Mexico); PAEC (Pakistan); SCSR (Poland); FCT (Portugal); JINR (Armenia, Belarus, Georgia, Ukraine, Uzbekistan); MST and MAE (Russia); MSTD (Serbia); MICINN and CPAN (Spain); Swiss Funding Agencies (Switzerland); NSC (Taipei); TUBITAK and TAEK (Turkey); STFC (United Kingdom); DOE and NSF (USA).

\bibliography{auto_generated}

\clearpage

\include{Tables}

\cleardoublepage\appendix\section{The CMS Collaboration \label{app:collab}}\begin{sloppypar}\hyphenpenalty=5000\widowpenalty=500\clubpenalty=5000\input{QCD-10-025-authorlist.tex}\end{sloppypar}
\end{document}

%% file: Tables.tex
\begin{table*}[htbH]
\begin{center}
\caption{Double-differential dijet mass cross section in the rapidity range $|y|_\text{max}<0.5$. The reference mass is the point at which the cross section is drawn in Figs.~\ref{fig:xsec} and ~\ref{fig:dataOverNLO} and is calculated as described in the text. The experimental systematic uncertainties of the individual dijet-mass bins are almost 100\% correlated. \label{tab:ybin0}}
\footnotesize
\begin{tabular}{|c|c|c|c|c|}
\hline
Mass Range & Reference Mass & Measured Cross Section & Statistical Uncertainty & Systematic Uncertainty \\
$(\!\TeV)$ & $(\!\TeV)$ & $(\text{pb}\!/\!\TeV)$ & \% & \% \\
\hline
\hline
 [0.156, 0.176] &      0.165 &                     $1.32\times 10^{6}$ &   $-$1.1, $+$1.1 &     $-10$, $+$11 \\

 [0.176, 0.197] &      0.186 &                     $7.26\times 10^{5}$ &   $-$1.4, $+$1.4 &     $-10$, $+$12 \\

 [0.197, 0.220] &      0.208 &                     $4.12\times 10^{5}$ &   $-$1.8, $+$1.8 &     $-11$, $+$12 \\

 [0.220, 0.244] &      0.231 &                     $2.35\times 10^{5}$ &   $-$0.7, $+$0.7 &     $-11$, $+$12 \\

 [0.244, 0.270] &      0.256 &                     $1.39\times 10^{5}$ &   $-$0.9, $+$0.9 &     $-11$, $+$12 \\

 [0.270, 0.296] &      0.282 &                     $8.18\times 10^{4}$ &   $-$0.7, $+$0.7 &     $-11$, $+$13 \\

 [0.296, 0.325] &      0.310 &                     $5.08\times 10^{4}$ &   $-$0.9, $+$0.9 &     $-11$, $+$13 \\

 [0.325, 0.354] &      0.339 &                     $3.18\times 10^{4}$ &   $-$1.1, $+$1.1 &     $-11$, $+$13 \\

 [0.354, 0.386] &      0.369 &                     $2.04\times 10^{4}$ &   $-$1.3, $+$1.3 &     $-12$, $+$13 \\

 [0.386, 0.419] &      0.402 &                     $1.28\times 10^{4}$ &   $-$1.1, $+$1.1 &     $-12$, $+$14 \\

 [0.419, 0.453] &      0.435 &                     $8.22\times 10^{3}$ &   $-$1.4, $+$1.4 &     $-12$, $+$14 \\

 [0.453, 0.489] &      0.470 &                     $5.42\times 10^{3}$ &   $-$1.6, $+$1.6 &     $-12$, $+$14 \\

 [0.489, 0.526] &      0.507 &                     $3.65\times 10^{3}$ &   $-$1.4, $+$1.5 &     $-13$, $+$14 \\

 [0.526, 0.565] &      0.545 &                     $2.44\times 10^{3}$ &   $-$1.7, $+$1.7 &     $-13$, $+$15 \\

 [0.565, 0.606] &      0.585 &                     $1.58\times 10^{3}$ &   $-$2.1, $+$2.1 &     $-13$, $+$15 \\

 [0.606, 0.649] &      0.627 &                     $1.05\times 10^{3}$ &   $-$2.5, $+$2.5 &     $-13$, $+$16 \\

 [0.649, 0.693] &      0.670 &                     $7.35\times 10^{2}$ &   $-$2.9, $+$3.0 &     $-14$, $+$16 \\

 [0.693, 0.740] &      0.716 &                     $5.05\times 10^{2}$ &   $-$3.4, $+$3.5 &     $-14$, $+$16 \\

 [0.740, 0.788] &      0.763 &                     $3.62\times 10^{2}$ &   $-$4.0, $+$4.2 &     $-14$, $+$17 \\

 [0.788, 0.838] &      0.812 &                     $2.45\times 10^{2}$ &   $-$4.8, $+$5.0 &     $-15$, $+$17 \\

 [0.838, 0.890] &      0.863 &                     $1.77\times 10^{2}$ &   $-$5.5, $+$5.8 &     $-15$, $+$18 \\

 [0.890, 0.944] &      0.916 &                     $1.13\times 10^{2}$ &   $-$6.8, $+$7.2 &     $-15$, $+$18 \\

 [0.944, 1.000] &      0.971 &                     $7.94\times 10^{1}$ &   $-$7.9, $+$8.5 &     $-16$, $+$19 \\

 [1.000, 1.118] &      1.055 &                     $4.76\times 10^{1}$ &   $-$7.0, $+$7.5 &     $-16$, $+$20 \\

 [1.118, 1.246] &      1.178 &                     $2.72\times 10^{1}$ &   $-$8.9, $+$9.8 &     $-17$, $+$21 \\

 [1.246, 1.383] &      1.310 &                     $1.14\times 10^{1}$ &   $-$13, $+$15 &     $-18$, $+$22 \\

 [1.383, 1.530] &      1.452 &                     $6.24$ &   $-$17, $+$21 &     $-19$, $+$24 \\

 [1.530, 1.687] &      1.604 &                     $2.48$ &   $-$26, $+$35 &     $-20$, $+$26 \\

 [1.687, 1.856] &      1.766 &                     $9.85\times 10^{-1}$ &   $-$40, $+$60 &     $-22$, $+$28 \\

 [1.856, 2.037] &      1.941 &                     $4.59\times 10^{-1}$ &   $-$54, $+$97 &     $-23$, $+$31 \\

 [2.037, 2.332] &      2.170 &                     $4.69\times 10^{-1}$ &   $-$43, $+$68 &     $-25$, $+$34 \\

 [2.332, 2.659] &      2.479 &                     $8.45\times 10^{-2}$ &   $-$83, $+$230 &     $-29$, $+$40 \\
\hline
\end{tabular}
\end{center}
\end{table*}

\begin{table*}[htbH]
\begin{center}
\caption{Double-differential dijet mass cross section in the rapidity range $0.5<|y|_\text{max}<1.0$. The reference mass is the point at which the cross section is drawn in Figs.~\ref{fig:xsec} and ~\ref{fig:dataOverNLO} and is calculated as described in the text. The experimental systematic uncertainties of the individual dijet-mass bins are almost 100\% correlated. \label{tab:ybin1}}
\footnotesize
\begin{tabular}{|c|c|c|c|c|}
\hline
Mass Range & Reference Mass & Measured Cross Section & Statistical Uncertainty & Systematic Uncertainty \\
$(\!\TeV)$ & $(\!\TeV)$ & $(\text{pb}\!/\!\TeV)$ & \% & \% \\
\hline
\hline
 [0.197, 0.220] &      0.208 &                     $1.74\times 10^{6}$ &   $-$0.8, $+$0.9 &     $-11$, $+$12 \\

 [0.220, 0.244] &      0.231 &                     $1.02\times 10^{6}$ &   $-$1.1, $+$1.1 &     $-11$, $+$12 \\

 [0.244, 0.270] &      0.256 &                     $6.00\times 10^{5}$ &   $-$1.4, $+$1.4 &     $-11$, $+$12 \\

 [0.270, 0.296] &      0.282 &                     $3.64\times 10^{5}$ &   $-$1.7, $+$1.8 &     $-11$, $+$12 \\

 [0.296, 0.325] &      0.310 &                     $2.22\times 10^{5}$ &   $-$0.7, $+$0.7 &     $-11$, $+$13 \\

 [0.325, 0.354] &      0.339 &                     $1.38\times 10^{5}$ &   $-$0.8, $+$0.9 &     $-11$, $+$13 \\

 [0.354, 0.386] &      0.369 &                     $8.64\times 10^{4}$ &   $-$1.0, $+$1.0 &     $-12$, $+$13 \\

 [0.386, 0.419] &      0.402 &                     $5.42\times 10^{4}$ &   $-$0.8, $+$0.8 &     $-12$, $+$13 \\

 [0.419, 0.453] &      0.435 &                     $3.55\times 10^{4}$ &   $-$1.0, $+$1.0 &     $-12$, $+$14 \\

 [0.453, 0.489] &      0.470 &                     $2.34\times 10^{4}$ &   $-$1.1, $+$1.2 &     $-12$, $+$14 \\

 [0.489, 0.526] &      0.507 &                     $1.53\times 10^{4}$ &   $-$0.9, $+$0.9 &     $-12$, $+$14 \\

 [0.526, 0.565] &      0.545 &                     $1.01\times 10^{4}$ &   $-$1.1, $+$1.1 &     $-13$, $+$15 \\

 [0.565, 0.606] &      0.585 &                     $6.90\times 10^{3}$ &   $-$1.3, $+$1.3 &     $-13$, $+$15 \\

 [0.606, 0.649] &      0.627 &                     $4.60\times 10^{3}$ &   $-$1.6, $+$1.6 &     $-13$, $+$15 \\

 [0.649, 0.693] &      0.670 &                     $3.15\times 10^{3}$ &   $-$1.4, $+$1.4 &     $-13$, $+$16 \\

 [0.693, 0.740] &      0.716 &                     $2.14\times 10^{3}$ &   $-$1.7, $+$1.7 &     $-14$, $+$16 \\

 [0.740, 0.788] &      0.763 &                     $1.48\times 10^{3}$ &   $-$2.0, $+$2.0 &     $-14$, $+$16 \\

 [0.788, 0.838] &      0.812 &                     $1.04\times 10^{3}$ &   $-$2.3, $+$2.4 &     $-14$, $+$17 \\

 [0.838, 0.890] &      0.863 &                     $7.42\times 10^{2}$ &   $-$2.7, $+$2.8 &     $-15$, $+$17 \\

 [0.890, 0.944] &      0.916 &                     $5.01\times 10^{2}$ &   $-$3.2, $+$3.3 &     $-15$, $+$18 \\

 [0.944, 1.000] &      0.971 &                     $3.37\times 10^{2}$ &   $-$3.8, $+$4.0 &     $-15$, $+$18 \\

 [1.000, 1.118] &      1.055 &                     $2.08\times 10^{2}$ &   $-$3.4, $+$3.5 &     $-16$, $+$19 \\

 [1.118, 1.246] &      1.178 &                     $9.94\times 10^{1}$ &   $-$4.7, $+$4.9 &     $-17$, $+$20 \\

 [1.246, 1.383] &      1.310 &                     $5.38\times 10^{1}$ &   $-$6.1, $+$6.5 &     $-18$, $+$22 \\

 [1.383, 1.530] &      1.452 &                     $2.73\times 10^{1}$ &   $-$8.3, $+$9.0 &     $-19$, $+$23 \\

 [1.530, 1.687] &      1.604 &                     $9.70$ &   $-$13, $+$15 &     $-20$, $+$25 \\

 [1.687, 1.856] &      1.766 &                     $5.73$ &   $-$17, $+$20 &     $-21$, $+$27 \\

 [1.856, 2.037] &      1.941 &                     $3.66$ &   $-$20, $+$25 &     $-23$, $+$30 \\

 [2.037, 2.332] &      2.170 &                     $1.12$ &   $-$28, $+$38 &     $-25$, $+$33 \\

 [2.332, 2.659] &      2.479 &                     $2.52\times 10^{-1}$ &   $-$54, $+$97 &     $-28$, $+$39 \\

 [2.659, 3.019] &      2.819 &                     $7.62\times 10^{-2}$ &   $-$83, $+$230 &     $-32$, $+$47 \\
\hline
\end{tabular}
\end{center}
\end{table*}

\begin{table*}[htbH]
\begin{center}
\caption{Double-differential dijet mass cross section in the rapidity range $1.0<|y|_\text{max}<1.5$. The reference mass is the point at which the cross section is drawn in Figs.~\ref{fig:xsec} and ~\ref{fig:dataOverNLO} and is calculated as described in the text. The experimental systematic uncertainties of the individual dijet-mass bins are almost 100\% correlated. \label{tab:ybin2}}
\footnotesize
\begin{tabular}{|c|c|c|c|c|}
\hline
Mass Range & Reference Mass & Measured Cross Section & Statistical Uncertainty & Systematic Uncertainty \\
$(\!\TeV)$ & $(\!\TeV)$ & $(\text{pb}\!/\!\TeV)$ & \% & \% \\
\hline
\hline
 [0.386, 0.419] &      0.402 &                     $1.84\times 10^{5}$ &   $-$2.2, $+$2.3 &     $-12$, $+$13 \\

 [0.419, 0.453] &      0.435 &                     $1.21\times 10^{5}$ &   $-$2.7, $+$2.8 &     $-12$, $+$13 \\

 [0.453, 0.489] &      0.470 &                     $7.77\times 10^{4}$ &   $-$3.3, $+$3.4 &     $-12$, $+$14 \\

 [0.489, 0.526] &      0.507 &                     $5.26\times 10^{4}$ &   $-$1.2, $+$1.2 &     $-12$, $+$14 \\

 [0.526, 0.565] &      0.545 &                     $3.56\times 10^{4}$ &   $-$1.5, $+$1.5 &     $-12$, $+$14 \\

 [0.565, 0.606] &      0.585 &                     $2.31\times 10^{4}$ &   $-$1.8, $+$1.8 &     $-13$, $+$15 \\

 [0.606, 0.649] &      0.627 &                     $1.60\times 10^{4}$ &   $-$2.1, $+$2.1 &     $-13$, $+$15 \\

 [0.649, 0.693] &      0.670 &                     $1.04\times 10^{4}$ &   $-$1.6, $+$1.6 &     $-13$, $+$15 \\

 [0.693, 0.740] &      0.716 &                     $7.20\times 10^{3}$ &   $-$1.8, $+$1.9 &     $-13$, $+$16 \\

 [0.740, 0.788] &      0.763 &                     $4.98\times 10^{3}$ &   $-$2.2, $+$2.2 &     $-14$, $+$16 \\

 [0.788, 0.838] &      0.812 &                     $3.35\times 10^{3}$ &   $-$2.6, $+$2.7 &     $-14$, $+$16 \\

 [0.838, 0.890] &      0.863 &                     $2.34\times 10^{3}$ &   $-$2.0, $+$2.1 &     $-14$, $+$17 \\

 [0.890, 0.944] &      0.916 &                     $1.65\times 10^{3}$ &   $-$2.4, $+$2.5 &     $-15$, $+$17 \\

 [0.944, 1.000] &      0.971 &                     $1.18\times 10^{3}$ &   $-$2.8, $+$2.9 &     $-15$, $+$18 \\

 [1.000, 1.118] &      1.055 &                     $6.61\times 10^{2}$ &   $-$2.6, $+$2.6 &     $-16$, $+$19 \\

 [1.118, 1.246] &      1.178 &                     $3.22\times 10^{2}$ &   $-$2.6, $+$2.7 &     $-16$, $+$20 \\

 [1.246, 1.383] &      1.310 &                     $1.57\times 10^{2}$ &   $-$3.6, $+$3.7 &     $-17$, $+$21 \\

 [1.383, 1.530] &      1.452 &                     $7.86\times 10^{1}$ &   $-$4.9, $+$5.1 &     $-18$, $+$22 \\

 [1.530, 1.687] &      1.603 &                     $3.80\times 10^{1}$ &   $-$6.8, $+$7.3 &     $-19$, $+$24 \\

 [1.687, 1.856] &      1.766 &                     $1.75\times 10^{1}$ &   $-$9.6, $+$11 &     $-20$, $+$26 \\

 [1.856, 2.037] &      1.941 &                     $8.32$ &   $-$13, $+$15 &     $-22$, $+$28 \\

 [2.037, 2.332] &      2.170 &                     $3.33$ &   $-$17, $+$20 &     $-23$, $+$31 \\

 [2.332, 2.659] &      2.478 &                     $1.83$ &   $-$21, $+$26 &     $-26$, $+$35 \\

 [2.659, 3.019] &      2.819 &                     $4.51\times 10^{-1}$ &   $-$40, $+$60 &     $-29$, $+$41 \\
\hline
\end{tabular}
\end{center}
\end{table*}

\begin{table*}[htbH]
\begin{center}
\caption{Double-differential dijet mass cross section in the rapidity range $1.5<|y|_\text{max}<2.0$. The reference mass is the point at which the cross section is drawn in Figs.~\ref{fig:xsec} and ~\ref{fig:dataOverNLO} and is calculated as described in the text. The experimental systematic uncertainties of the individual dijet-mass bins are almost 100\% correlated. \label{tab:ybin3}}
\footnotesize
\begin{tabular}{|c|c|c|c|c|}
\hline
Mass Range & Reference Mass & Measured Cross Section & Statistical Uncertainty & Systematic Uncertainty \\
$(\!\TeV)$ & $(\!\TeV)$ & $(\text{pb}\!/\!\TeV)$ & \% & \% \\
\hline
\hline
 [0.565, 0.606] &      0.585 &                     $6.68\times 10^{4}$ &   $-$3.1, $+$3.2 &     $-12$, $+$14 \\

 [0.606, 0.649] &      0.627 &                     $4.52\times 10^{4}$ &   $-$3.7, $+$3.9 &     $-12$, $+$14 \\

 [0.649, 0.693] &      0.670 &                     $3.05\times 10^{4}$ &   $-$4.5, $+$4.7 &     $-12$, $+$14 \\

 [0.693, 0.740] &      0.716 &                     $2.02\times 10^{4}$ &   $-$1.7, $+$1.7 &     $-13$, $+$15 \\

 [0.740, 0.788] &      0.763 &                     $1.47\times 10^{4}$ &   $-$2.0, $+$2.0 &     $-13$, $+$15 \\

 [0.788, 0.838] &      0.812 &                     $1.04\times 10^{4}$ &   $-$2.3, $+$2.4 &     $-13$, $+$15 \\

 [0.838, 0.890] &      0.863 &                     $6.92\times 10^{3}$ &   $-$2.8, $+$2.8 &     $-13$, $+$16 \\

 [0.890, 0.944] &      0.916 &                     $4.77\times 10^{3}$ &   $-$2.1, $+$2.1 &     $-14$, $+$16 \\

 [0.944, 1.000] &      0.971 &                     $3.41\times 10^{3}$ &   $-$2.4, $+$2.4 &     $-14$, $+$16 \\

 [1.000, 1.118] &      1.055 &                     $2.04\times 10^{3}$ &   $-$2.1, $+$2.2 &     $-14$, $+$17 \\

 [1.118, 1.246] &      1.178 &                     $1.04\times 10^{3}$ &   $-$2.0, $+$2.0 &     $-15$, $+$18 \\

 [1.246, 1.383] &      1.310 &                     $5.20\times 10^{2}$ &   $-$2.7, $+$2.7 &     $-16$, $+$19 \\

 [1.383, 1.530] &      1.452 &                     $2.60\times 10^{2}$ &   $-$3.6, $+$3.8 &     $-17$, $+$21 \\

 [1.530, 1.687] &      1.604 &                     $1.45\times 10^{2}$ &   $-$4.7, $+$5.0 &     $-18$, $+$22 \\

 [1.687, 1.856] &      1.766 &                     $6.31\times 10^{1}$ &   $-$5.1, $+$5.3 &     $-19$, $+$24 \\

 [1.856, 2.037] &      1.941 &                     $3.24\times 10^{1}$ &   $-$6.8, $+$7.3 &     $-21$, $+$26 \\

 [2.037, 2.332] &      2.170 &                     $1.18\times 10^{1}$ &   $-$8.9, $+$9.7 &     $-23$, $+$29 \\

 [2.332, 2.659] &      2.479 &                     $4.37$ &   $-$14, $+$16 &     $-26$, $+$34 \\

 [2.659, 3.019] &      2.820 &                     $1.52$ &   $-$22, $+$28 &     $-29$, $+$41 \\

 [3.019, 3.854] &      3.344 &                     $1.31\times 10^{-1}$ &   $-$48, $+$79 &     $-35$, $+$54 \\
\hline
\end{tabular}
\end{center}
\end{table*}

\begin{table*}[htbH]
\begin{center}
\caption{Double-differential dijet mass cross section in the rapidity range $2.0<|y|_\text{max}<2.5$. The reference mass is the point at which the cross section is drawn in Figs.~\ref{fig:xsec} and ~\ref{fig:dataOverNLO} and is calculated as described in the text. The experimental systematic uncertainties of the individual dijet-mass bins are almost 100\% correlated. \label{tab:ybin4}}
\footnotesize
\begin{tabular}{|c|c|c|c|c|}
\hline
Mass Range & Reference Mass & Measured Cross Section & Statistical Uncertainty & Systematic Uncertainty \\
$(\!\TeV)$ & $(\!\TeV)$ & $(\text{pb}\!/\!\TeV)$ & \% & \% \\
\hline
\hline
 [0.649, 0.693] &      0.670 &                     $1.00\times 10^{5}$ &   $-$2.5, $+$2.5 &     $-13$, $+$15 \\

 [0.693, 0.740] &      0.716 &                     $6.70\times 10^{4}$ &   $-$2.9, $+$3.0 &     $-13$, $+$15 \\

 [0.740, 0.788] &      0.763 &                     $4.63\times 10^{4}$ &   $-$3.5, $+$3.6 &     $-13$, $+$15 \\

 [0.788, 0.838] &      0.812 &                     $3.29\times 10^{4}$ &   $-$4.1, $+$4.2 &     $-13$, $+$15 \\

 [0.838, 0.890] &      0.863 &                     $2.31\times 10^{4}$ &   $-$4.8, $+$5.0 &     $-13$, $+$16 \\

 [0.890, 0.944] &      0.916 &                     $1.52\times 10^{4}$ &   $-$1.8, $+$1.9 &     $-14$, $+$16 \\

 [0.944, 1.000] &      0.971 &                     $1.11\times 10^{4}$ &   $-$2.1, $+$2.2 &     $-14$, $+$16 \\

 [1.000, 1.118] &      1.055 &                     $6.41\times 10^{3}$ &   $-$1.9, $+$1.9 &     $-14$, $+$16 \\

 [1.118, 1.246] &      1.178 &                     $3.26\times 10^{3}$ &   $-$2.6, $+$2.6 &     $-14$, $+$17 \\

 [1.246, 1.383] &      1.310 &                     $1.59\times 10^{3}$ &   $-$2.2, $+$2.3 &     $-15$, $+$18 \\

 [1.383, 1.530] &      1.452 &                     $8.39\times 10^{2}$ &   $-$3.0, $+$3.1 &     $-16$, $+$18 \\

 [1.530, 1.687] &      1.604 &                     $4.01\times 10^{2}$ &   $-$4.2, $+$4.4 &     $-16$, $+$19 \\

 [1.687, 1.856] &      1.766 &                     $1.80\times 10^{2}$ &   $-$4.1, $+$4.2 &     $-17$, $+$21 \\

 [1.856, 2.037] &      1.941 &                     $8.96\times 10^{1}$ &   $-$5.6, $+$5.9 &     $-18$, $+$22 \\

 [2.037, 2.332] &      2.170 &                     $3.75\times 10^{1}$ &   $-$6.8, $+$7.2 &     $-19$, $+$24 \\

 [2.332, 2.659] &      2.479 &                     $9.44$ &   $-$9.4, $+$10 &     $-22$, $+$27 \\

 [2.659, 3.019] &      2.819 &                     $3.52$ &   $-$15, $+$17 &     $-25$, $+$32 \\

 [3.019, 3.854] &      3.338 &                     $3.29\times 10^{-1}$ &   $-$31, $+$43 &     $-30$, $+$41 \\
\hline
\end{tabular}
\end{center}
\end{table*}

%% file: QCD-10-025-authorlist.tex
\textbf{Yerevan Physics Institute,  Yerevan,  Armenia}\\*[0pt]
S.~Chatrchyan, V.~Khachatryan, A.M.~Sirunyan, A.~Tumasyan
\vskip\cmsinstskip
\textbf{Institut f\"{u}r Hochenergiephysik der OeAW,  Wien,  Austria}\\*[0pt]
W.~Adam, T.~Bergauer, M.~Dragicevic, J.~Er\"{o}, C.~Fabjan, M.~Friedl, R.~Fr\"{u}hwirth, V.M.~Ghete, J.~Hammer\cmsAuthorMark{1}, S.~H\"{a}nsel, M.~Hoch, N.~H\"{o}rmann, J.~Hrubec, M.~Jeitler, G.~Kasieczka, W.~Kiesenhofer, M.~Krammer, D.~Liko, I.~Mikulec, M.~Pernicka, H.~Rohringer, R.~Sch\"{o}fbeck, J.~Strauss, F.~Teischinger, P.~Wagner, W.~Waltenberger, G.~Walzel, E.~Widl, C.-E.~Wulz
% ...
\vskip\cmsinstskip
% ...
1:~~Also at CERN, European Organization for Nuclear Research, Geneva, Switzerland\\
% ...


%% file: QCD-10-025_temp.bbl
\providecommand{\href}[2]{#2}\begingroup\raggedright\begin{thebibliography}{10}%
\makeatletter
\providecommand{\hrefCMSnoop }[0]{\@secondoftwo}%
\makeatother

\bibitem{D0}
\hrefCMSnoop {} {{ D0} Collaboration, ``Measurement of the Dijet Invariant Mass
  Cross Section in $p\overline{p}$ Collisions at $\sqrt{s} = 1.96{\TeV}$'',}
  \textit{ Phys. Lett.} \textbf{ B693} (2010) 531.
  \href{http://dx.doi.org/10.1016/j.physletb.2010.09.013}{\texttt{
  doi:10.1016/j.physletb.2010.09.013}}.

\bibitem{CDF}
\hrefCMSnoop {} {{ CDF} Collaboration, ``Search for new particles decaying into
  dijets in proton-antiproton collisions at $\sqrt{s} = 1.96{\TeV}$'',}
  \textit{ Phys. Rev.} \textbf{ D79} (2009) 112002.
  \href{http://dx.doi.org/10.1103/PhysRevD.79.112002}{\texttt{
  doi:10.1103/PhysRevD.79.112002}}.

\bibitem{ATLAS}
\hrefCMSnoop {} {{ ATLAS} Collaboration, ``Measurement of inclusive jet and
  dijet cross sections in proton-proton collisions at {$7\TeV$} centre-of-mass
  energy with the {ATLAS} detector'',} \textit{ Eur. Phys. J.} \textbf{ C71}
  (2011) 1512. \href{http://dx.doi.org/10.1140/epjc/s10052-010-1512-2}{\texttt{
  doi:10.1140/epjc/s10052-010-1512-2}}.

\bibitem{CMSSearchMass}
\hrefCMSnoop {} {{ CMS} Collaboration, ``Search for Dijet Resonances in 7 {TeV}
  $pp$ Collisions at {CMS}'',} \textit{ Phys. Rev. Lett.} \textbf{ 105} (2010)
  211801. \href{http://dx.doi.org/10.1103/PhysRevLett.105.211801}{\texttt{
  doi:10.1103/PhysRevLett.105.211801}}.

\bibitem{CMSSearchRatio}
\hrefCMSnoop {} {{ CMS} Collaboration, ``Search for Quark Compositeness with
  the Dijet Centrality Ratio in $pp$ Collisions at $\sqrt{s}=7{\TeV}$'',}
  \textit{ Phys. Rev. Lett.} \textbf{ 105} (2010) 262001.
  \href{http://dx.doi.org/10.1103/PhysRevLett.105.262001}{\texttt{
  doi:10.1103/PhysRevLett.105.262001}}.

\bibitem{CMSSearchAngular}
\hrefCMSnoop {} {{ CMS} Collaboration, ``Measurement of Dijet Angular
  Distributions and Search for Quark Compositeness in $pp$ Collisions at 7
  {TeV}'',} (2010). \href{http://www.arXiv.org/abs/1102.2020v1}{\texttt{
  arXiv:1102.2020v1}}. Submitted to \textit{Physical Review Letters}.

\bibitem{CMS}
\hrefCMSnoop {} {{ CMS} Collaboration, ``The {CMS} experiment at the {CERN}
  {LHC}'',} \textit{ JINST} \textbf{ 3} (2008) S08004.
  \href{http://dx.doi.org/10.1088/1748-0221/3/08/S08004}{\texttt{
  doi:10.1088/1748-0221/3/08/S08004}}.

\bibitem{AKT}
\hrefCMSnoop {} {M.~Cacciari, G.~P. Salam, and G.~Soyez, ``The anti-kt jet
  clustering algorithm'',} \textit{ JHEP} \textbf{ 04} (2008) 063.
  \href{http://dx.doi.org/10.1088/1126-6708/2008/04/063}{\texttt{
  doi:10.1088/1126-6708/2008/04/063}}.

\bibitem{PFLOW}
\href {http://cdsweb.cern.ch/record/1194487} {{ CMS} Collaboration,
  ``Particle--Flow Event Reconstruction in {CMS} and Performance for Jets,
  Taus, and {\MET}'',} \textit{ CMS Physics Analysis Summary} \textbf{
  CMS-PAS-PFT-09-001} (2009).

\bibitem{PYTHIA}
\hrefCMSnoop {} {T.~Sj{\"o}strand, S.~Mrenna, and P.~Skands, ``{\PYTHIA} 6.4
  physics and manual'',} \textit{ JHEP} \textbf{ 05} (2006) 026.
  \href{http://dx.doi.org/10.1088/1126-6708/2006/05/026}{\texttt{
  doi:10.1088/1126-6708/2006/05/026}}.

\bibitem{GEANT4}
\hrefCMSnoop {} {S.~Agostinelli {et~al.}, ``Geant 4 -- A Simulation Toolkit'',}
  \textit{ Nucl. Inst. Meth.} \textbf{ A506} (2003) 250.
  \href{http://dx.doi.org/10.1016/S0168-9002(03)01368-8}{\texttt{
  doi:10.1016/S0168-9002(03)01368-8}}.

\bibitem{JME-10-010}
\href {http://cdsweb.cern.ch/record/1308178} {{ CMS} Collaboration,
  ``Determination of the Jet Energy Scale in {CMS} with pp Collisions at
  $\sqrt{s}= 7$ {TeV}'',} \textit{ CMS Physics Analysis Summary} \textbf{
  CMS-PAS-JME-10-010} (2010).

\bibitem{HLT}
\hrefCMSnoop {} {{ CMS} Collaboration, ``The {CMS} High Level Trigger'',}
  \textit{ Eur. Phys. J.} \textbf{ C46} (2006) 605.
  \href{http://dx.doi.org/10.1140/epjc/s2006-02495-8}{\texttt{
  doi:10.1140/epjc/s2006-02495-8}}.

\bibitem{TRK-10-005}
\href {http://cdsweb.cern.ch/record/1279383} {{ CMS} Collaboration, ``Tracking
  and Primary Vertex Results in First 7 {TeV} Collisions'',} \textit{ CMS
  Physics Analysis Summary} \textbf{ CMS-PAS-TRK-10-005} (2010).

\bibitem{POINT}
\hrefCMSnoop {} {G.~D.Lafferty and T.~R.Wyatt, ``Where to stick your data
  points: The treatment of measurements within wide bins'',} \textit{ Nucl.
  Inst. Meth.} \textbf{ A355} (1995) 541.
  \href{http://dx.doi.org/10.1016/0168-9002(94)01112-5}{\texttt{
  doi:10.1016/0168-9002(94)01112-5}}.

\bibitem{LUMI}
\href {http://cdsweb.cern.ch/record/1328359} {{ CMS} Collaboration, ``Absolute
  luminosity normalization'',} \textit{ CERN Detector Performance Studies}
  \textbf{ CMS-DP-2011-002} (2011).

\bibitem{JME-10-014}
\href {http://cdsweb.cern.ch/record/1339945} {{ CMS} Collaboration, ``Jet
  Energy Resolution in CMS at $\sqrt{s}=7$ {TeV}'',} \textit{ CMS Physics
  Analysis Summary} \textbf{ CMS-PAS-JME-10-014} (2010).

\bibitem{NLO}
\hrefCMSnoop {} {Z.~Nagy, ``Next-to-leading order calculation of three jet
  observables in hadron-hadron collision'',} \textit{ Phys. Rev.} \textbf{ D68}
  (2003) 094002. \href{http://dx.doi.org/10.1103/PhysRevD.68.094002}{\texttt{
  doi:10.1103/PhysRevD.68.094002}}.

\bibitem{fastNLO}
\hrefCMSnoop {} {T.~Kluge, K.~Rabbertz, and M.~Wobisch, ``{fastNLO}: Fast
  {pQCD} calculations for {PDF} fits'',}
  \href{http://www.arXiv.org/abs/hep-ph/0609285v2}{\texttt{
  arXiv:hep-ph/0609285v2}}.

\bibitem{CT10}
\hrefCMSnoop {} {H.-L. Lai {et~al.}, ``New parton distributions for collider
  physics'',} \textit{ Phys. Rev.} \textbf{ D82} (2010) 074024.
  \href{http://dx.doi.org/10.1103/PhysRevD.82.074024}{\texttt{
  doi:10.1103/PhysRevD.82.074024}}.

\bibitem{MSTW}
\hrefCMSnoop {} {A.~D. Martin {et~al.}, ``Parton distributions for the
  {LHC}'',} \textit{ Eur. Phys. J.} \textbf{ C63} (2009) 189.
  \href{http://dx.doi.org/10.1140/epjc/s10052-009-1072-5}{\texttt{
  doi:10.1140/epjc/s10052-009-1072-5}}.

\bibitem{NNPDF}
\hrefCMSnoop {} {R.~D. Ball {et~al.}, ``A first unbiased global {NLO}
  determination of parton distributions and their uncertainties'',} \textit{
  Nucl. Phys.} \textbf{ B838} (2010) 136.
  \href{http://dx.doi.org/10.1016/j.nuclphysb.2010.05.008}{\texttt{
  doi:10.1016/j.nuclphysb.2010.05.008}}.

\bibitem{PDF4LHC}
\hrefCMSnoop {} {S.~Alekhin {et~al.}, ``The {PDF4LHC} Working Group Interim
  Report.'',} (2011). \href{http://www.arXiv.org/abs/1101.0536v1}{\texttt{
  arXiv:1101.0536v1}}.

\bibitem{D6T}
\hrefCMSnoop {} {R.~Field, ``Early {LHC} Underlying Event Data-Findings and
  Surprises'',} (2010). \href{http://www.arXiv.org/abs/1010.3558v1}{\texttt{
  arXiv:1010.3558v1}}.

\bibitem{Z2}
The {\sc pythia6} {\sc z2} tune is identical to the {\sc z1} tune described
  in~\cite{D6T} except that {\sc z2} uses the CTEQ6L PDF while {\sc z1} uses
  CTEQ5L.

\bibitem{HERWIG}
\hrefCMSnoop {} {M.~B{\"a}hr {et~al.}, ``{Herwig++} Physics and Manual'',}
  \textit{ Eur. Phys. J.} \textbf{ C58} (2008) 639.
  \href{http://dx.doi.org/10.1140/epjc/s10052-008-0798-9}{\texttt{
  doi:10.1140/epjc/s10052-008-0798-9}}.

\end{thebibliography}\endgroup
